\documentclass[aip,graphicx,amsmath,amssymb,reprint]{revtex4-2}

\usepackage{epsfig}
\usepackage{epstopdf}
\usepackage{graphicx}
\usepackage{float}
\usepackage{dcolumn}
\usepackage{bm}
\usepackage[utf8]{inputenc}
\usepackage[T1]{fontenc}
\usepackage{mathptmx}
\usepackage{etoolbox}

\draft 

\begin{document}


\title{Miniaturized time-correlated single-photon counting module for time-of-flight non-line-of-sight imaging applications} 



\author{Jie Wu}
\affiliation{Hefei National Research Center for Physical Sciences at the Microscale and School of Physical Sciences, University of Science and Technology of China, Hefei 230026, China}
\affiliation{CAS Center for Excellence in Quantum Information and Quantum Physics, University of Science and Technology of China, Hefei 230026, China}

\author{Chao Yu}
\email[]{yuch@ustc.edu.cn}
\affiliation{Hefei National Research Center for Physical Sciences at the Microscale and School of Physical Sciences, University of Science and Technology of China, Hefei 230026, China}
\affiliation{CAS Center for Excellence in Quantum Information and Quantum Physics, University of Science and Technology of China, Hefei 230026, China}

\author{Jian-Wei Zeng}
\affiliation{Hefei National Research Center for Physical Sciences at the Microscale and School of Physical Sciences, University of Science and Technology of China, Hefei 230026, China}
\affiliation{CAS Center for Excellence in Quantum Information and Quantum Physics, University of Science and Technology of China, Hefei 230026, China}

\author{Chen Dai}
\affiliation{Hefei National Research Center for Physical Sciences at the Microscale and School of Physical Sciences, University of Science and Technology of China, Hefei 230026, China}
\affiliation{CAS Center for Excellence in Quantum Information and Quantum Physics, University of Science and Technology of China, Hefei 230026, China}

\author{Feihu Xu}
\affiliation{Hefei National Research Center for Physical Sciences at the Microscale and School of Physical Sciences, University of Science and Technology of China, Hefei 230026, China}
\affiliation{CAS Center for Excellence in Quantum Information and Quantum Physics, University of Science and Technology of China, Hefei 230026, China}
\affiliation{Hefei National Laboratory, University of Science and Technology of China, Hefei 230088, China}

\author{Jun Zhang}
\email[]{zhangjun@ustc.edu.cn}
\affiliation{Hefei National Research Center for Physical Sciences at the Microscale and School of Physical Sciences, University of Science and Technology of China, Hefei 230026, China}
\affiliation{CAS Center for Excellence in Quantum Information and Quantum Physics, University of Science and Technology of China, Hefei 230026, China}
\affiliation{Hefei National Laboratory, University of Science and Technology of China, Hefei 230088, China}


\date{\today}

\begin{abstract}
Single-photon time-of-flight (TOF) non-line-of-sight (NLOS) imaging enables the high-resolution reconstruction of objects outside the field of view. The compactness of TOF NLOS imaging systems, entailing the miniaturization of key components within such systems is crucial for practical applications. Here, we present a miniaturized four-channel time-correlated single-photon counting module dedicated to TOF NLOS imaging applications. The module achieves excellent performance with a 10 ps bin size and 27.4 ps minimum root-mean-square time resolution. We present the results of TOF NLOS imaging experiment using an InGaAs/InP single-photon detector and the time-correlated single-photon counting module, and show that a 6.3 cm lateral resolution and 2.3 cm depth resolution can be achieved under the conditions of 5 m imaging distance and 1 ms pixel dwell time.
\end{abstract}


\maketitle 
\makeatletter
\newcommand{\rmnum}[1]{\romannumeral#1}
\newcommand{\Rmnum}[1]{\expandafter\@slowromancap\romannumeral#1@}
\makeatother

\section{Introduction}

Non-line-of-sight (NLOS) imaging~\cite{KHD09,AMP18,FVW20} is a novel technique for reconstructing the details of obscured scenes from indirect light that has been  scattered multiple times. The ability to ``see through walls'' has attracted extensive interest due to its potential applications in medical detection, robotic surgery, earthquake relief, and autonomous driving. Over the past decade, many studies have attempted to improve the performance of NLOS imaging techniques in terms of resolution, distance, and imaging rate~\cite{VWG12,K14,MJA15,WLH21,WZH21,NBB21}. Currently, the main implementation schemes for NLOS imaging include time-of-flight (TOF) NLOS imaging~\cite{VWG12}, coherence-based NLOS imaging~\cite{MHR20}, and intensity-based NLOS imaging~\cite{KPM16}. In practice, TOF NLOS imaging systems are widely used because of their robustness to the environment, low cost, and high resolution.

Generally, the echo signal of a TOF NLOS imaging system is extremely weak because of the multiple scattering. Since single-photon detector (SPD) provides the ultimate sensitivity for photon detection, using SPD instead of conventional photon detector can greatly improve the imaging sensitivity and imaging distance of TOF NLOS imaging systems. Recently, TOF NLOS imaging experiments using SPDs have been conducted for realistic applications~\cite{WLH21,WZH21}. In practical scenarios, the single-photon TOF NLOS imaging system will be integrated as a vehicle-mounted or handheld device, so the use of miniaturized components is crucial. Among various SPD candidates, semiconductor SPDs based on single-photon avalanche diodes have the advantages of small size, low cost, and ease-of-use~\cite{Z15}. Recent studies have reported compact semiconductor SPDs in different bands, including ultraviolet~\cite{YLZ23,YYL23}, visible~\cite{FL20,CA19}, and near-infrared~\cite{YSX17,JGF18,XY23}, which could be highly suitable for single-photon TOF NLOS imaging. Therefore, apart from the SPD module, it is also necessary to miniaturize the remaining components, such as the time measurement module, in single-photon TOF NLOS imaging systems.

In single-photon TOF NLOS imaging applications, the arrival time of an echo signal is usually measured by a time-correlated single-photon counting (TCSPC) module. Developing a highly integrated TCSPC module is crucial for practical single-photon TOF NLOS imaging systems. There are several approaches to achieve high-precision TCSPC. Conventionally, using a time-to-amplitude convertor provides an analogue method of realizing picosecond time resolution~\cite{BBW01}. However, this method has inherent disadvantages, such as environmental sensitivity, a short measurement range, and a relatively low count rate. In contrast, digital methods, i.e., a time-to-digital converter (TDC), based on the delay line architecture in field-programmable gate arrays (FPGAs) are widely used due to the advantages of high resolution and flexible design~\cite{JZ08,WZS19,PPC22,CXD18,YXL22}. Digital methods can also be implemented in application-specific integrated circuits (ASICs)~\cite{DSH00,CLH14}. Currently, high-performance universal TDC ASIC chips are commercially available, which greatly helps the development of an integrated TCSPC system.

\begin{figure*}[htbp]
    \includegraphics[width=16 cm]{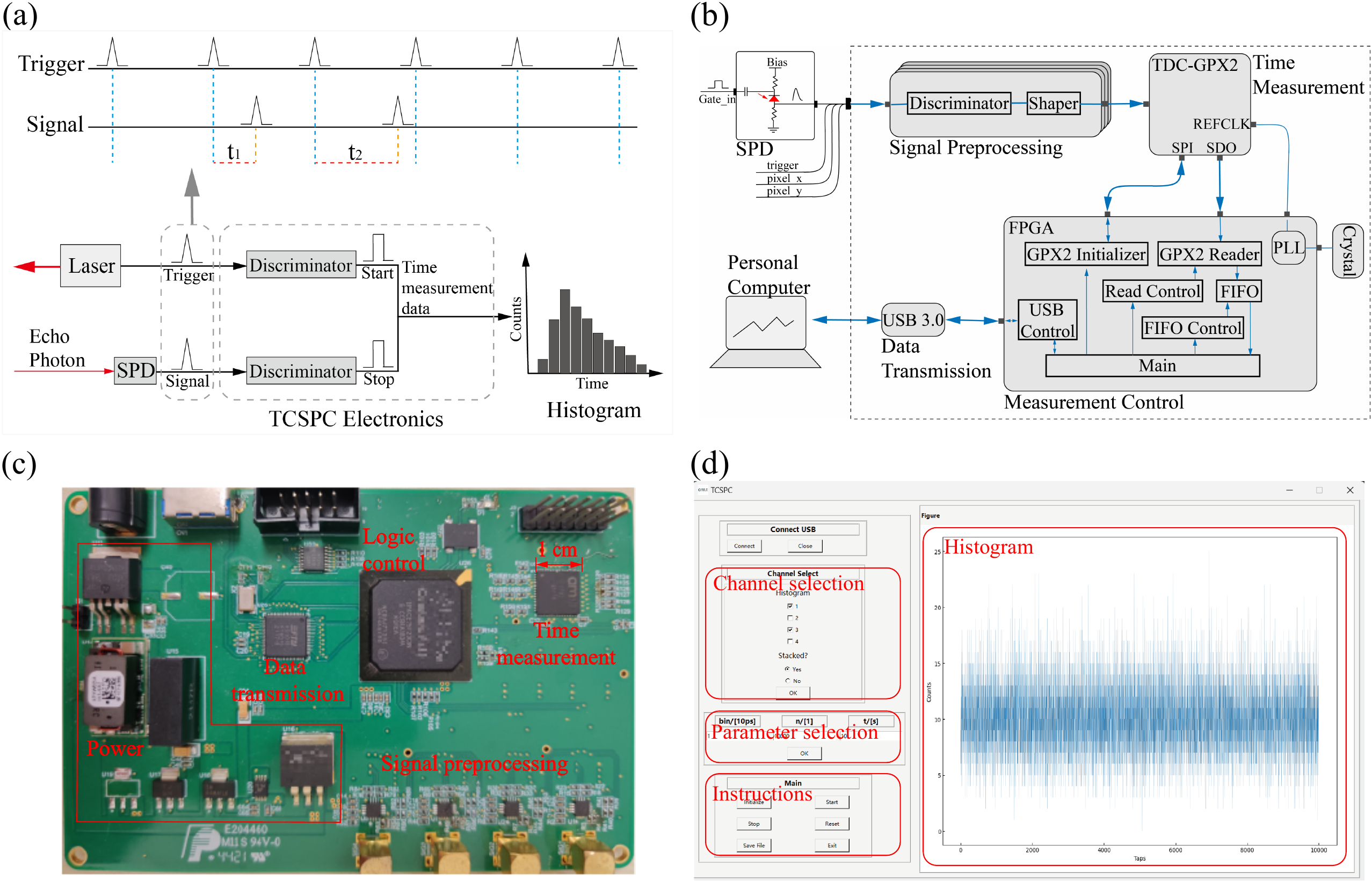}
    \caption{(a) The principle of TCSPC technique. (b) The schematic diagram of the whole application system with TCSPC, including a SPD, a TCSPC module and a personal computer. (c) The photograph of the TCSPC testing board. (d) The software interface for the TCSPC module.}
    \label{fig1}
\end{figure*}

In this paper, we present a miniaturized four-channel TCSPC module based on a commercial TDC ASIC chip, i.e., TDC-GPX2 (Sciosense)~\cite{GPX2}. An FPGA is used to control the chip and handle the time measurement processing. A high-speed universal serial bus (USB) 3.0 interface is used to transmit the time tag data to a personal computer. A user-friendly software application allows time tag data to be converted into a histogram. After characterizing the primary parameters, including the timing resolution and non-linearity, we perform an NLOS imaging experiment using the TCSPC module. The experimental results show that the TCSPC module is suitable for the implementation of a compact single-photon TOF NLOS imaging system.

\section{TCSPC MODULE DESIGN AND CHARACTERIZATION}

The principle of the TCSPC technique is shown in Fig.~\ref{fig1} (a). A pulsed laser illuminates the target with high repetition frequency. The echo photons are collected and detected by an SPD, and the TCSPC electronics measure the precise time difference between the photon arrival time and the reference laser pulse. The time measurement data are then aggregated to form a histogram, showing the intensity distribution of the echo photons at a given time.

The whole application scheme using the TCSPC technique is depicted in Fig.~\ref{fig1} (b). The system consists of an SPD, TCSPC module and personal computer. In the experiment, the SPD is implemented using a high-performance InGaAs/InP single-photon avalanche diode~\cite{FC20}. To minimize the effect of stray photons, the SPD operates in gated mode. The input gate clocks are synchronized to the laser pulses, and the gate repetition frequency and gate width are set to 10 MHz and 10 ns, respectively. Following the standard calibration procedure~\cite{Z15}, the SPD exhibits excellent performance of 54\% photon detection efficiency at 1550 nm, 2000 cps dark count rate, and 110 ps timing jitter. Moreover, the delay time between the gates and the laser pulses can be tuned for each pixel during the scanning process in the TOF NLOS imaging experiment.

The TCSPC module includes signal preprocessing, time measurement, logic control, and data transmission functions, as shown in Fig.~\ref{fig1} (b). The module supports input signals across four channels. In the TOF NLOS imaging experiment, the four input signals are the start signal synchronized to the laser pulses, SPD output signal, and two-dimensional pixel scan signals (pixel $x$ and pixel $y$). The input signals are first discriminated, and then converted into standard low voltage differential signaling (LVDS) with a fixed pulse width of 10 ns using shapers. The LVDS signals are fed into TDC-GPX2 for time measurement. An FPGA is used to initialize the TDC chip and read out the time tag data, either in single-data-rate mode or in double-data-rate mode. The FPGA also writes instructions to the chip registers via a serial peripheral interface (SPI), allowing the operation modes of the TCSPC module to be configured. The TCSPC module is implemented on an 8 cm $\times$ 12 cm six-layer printed circuit board for testing purposes, in which the size of time measurement component is $\sim$1 cm $\times$ 1 cm, as shown in Fig.~\ref{fig1} (c).

When the module is running, the arrival time of the input signal in each channel is tagged based on the reference clocks generated by an external high-accuracy crystal and a phase-locked loop (PLL) inside the FPGA. The bin size is tuned by dividing the reference clocks with a configurable factor. The time tag data consist of a coarse time package and a fine time package. The coarse time is measured by counting the reference clock periods, while the fine time is measured by counting the divided clock periods. During the experiment, the frequency of the reference clock is set to 10 MHz. Considering the total timing jitter of the NLOS imaging system is approximately 100 ps, the division factor is initialized to 10000, corresponding to a 10 ps bin size. The length of the fine time package is set to 14 bits, while the length of the coarse time package is customized according to the repetition frequency of the NLOS imaging system.

On the arrival of the input signals, the created time tag data are immediately sent to the FPGA. To avoid data stacking, a memory architecture of first in first out (FIFO) is designed inside the FPGA for data caching. In the TCSPC module, input signals from channel 1 are defined as start signals, and input signals from other channels are defined as stop signals. The time intervals between the start and stop signals are calculated in real-time. The data are then encoded with channel information, and transmitted to the personal computer via the USB3.0 interface for data postprocessing.
On the personal computer, a user-friendly software is developed using PyCharm, as shown in Fig.~\ref{fig1} (d).
The software can flexibly configure various TCSPC parameters, including the active channels, bin size, histogram depth, and measurement time, and also periodically converts the time tag data to a histogram.

\begin{figure}[htbp]
    \includegraphics[width=8.5 cm]{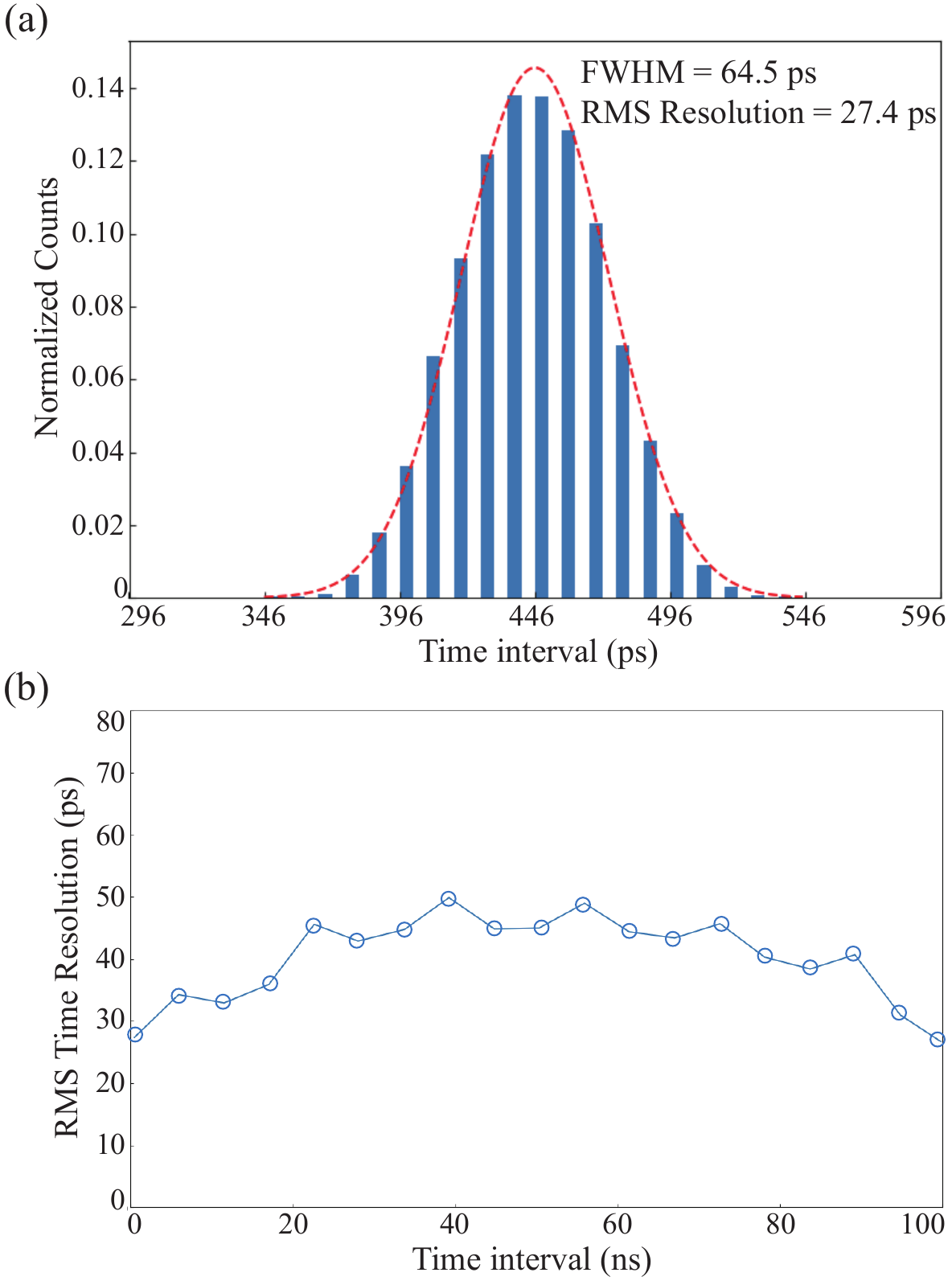}
    \caption{(a) Typical measured histogram of the time interval in the cable delay test. The bin size is configured as 10 ps. (b) RMS time resolution measurements by varying the cable delays.}
    \label{fig2}
\end{figure}

We then characterize the time resolution and non-linearity of the TCSPC module. The time resolution is evaluated using the cable delay test technique~\cite{SAL06}. Pair signals under test are generated from the same clock signal with different cable delays, and connected to channel 1 and channel 2 of the TCSPC module. Figure.~\ref{fig2} (a) plots a typical time-interval histogram between paired signals, which follows a Gaussian distribution. Given that the total number of counts in the measurement is of order 1 million, the root-mean-square (RMS) time resolution is 27.4 ps, as calculated by
\begin{equation}
    \label{RMS}
    t_{RMS} = \sqrt{\frac{\sum_{i=1}^{n}{(t_i-t_{mean})}^2}{n-1}}.
\end{equation}
The measurements are repeated for various cable delays of up to 100 ns. As shown in Fig.~\ref{fig2} (b), the RMS time resolution fluctuates from $\sim$27 ps to $\sim$50 ps. In the NLOS imaging applications, the full width at half maximum (FWHM) time resolution is usually used for calculating the spatial resolution. By fitting the Gaussian distribution, the minimum FWHM time resolution of the TCSPC module is calculated as 64.5 ps.

\begin{figure}[htbp]
    \includegraphics[width=8.5 cm]{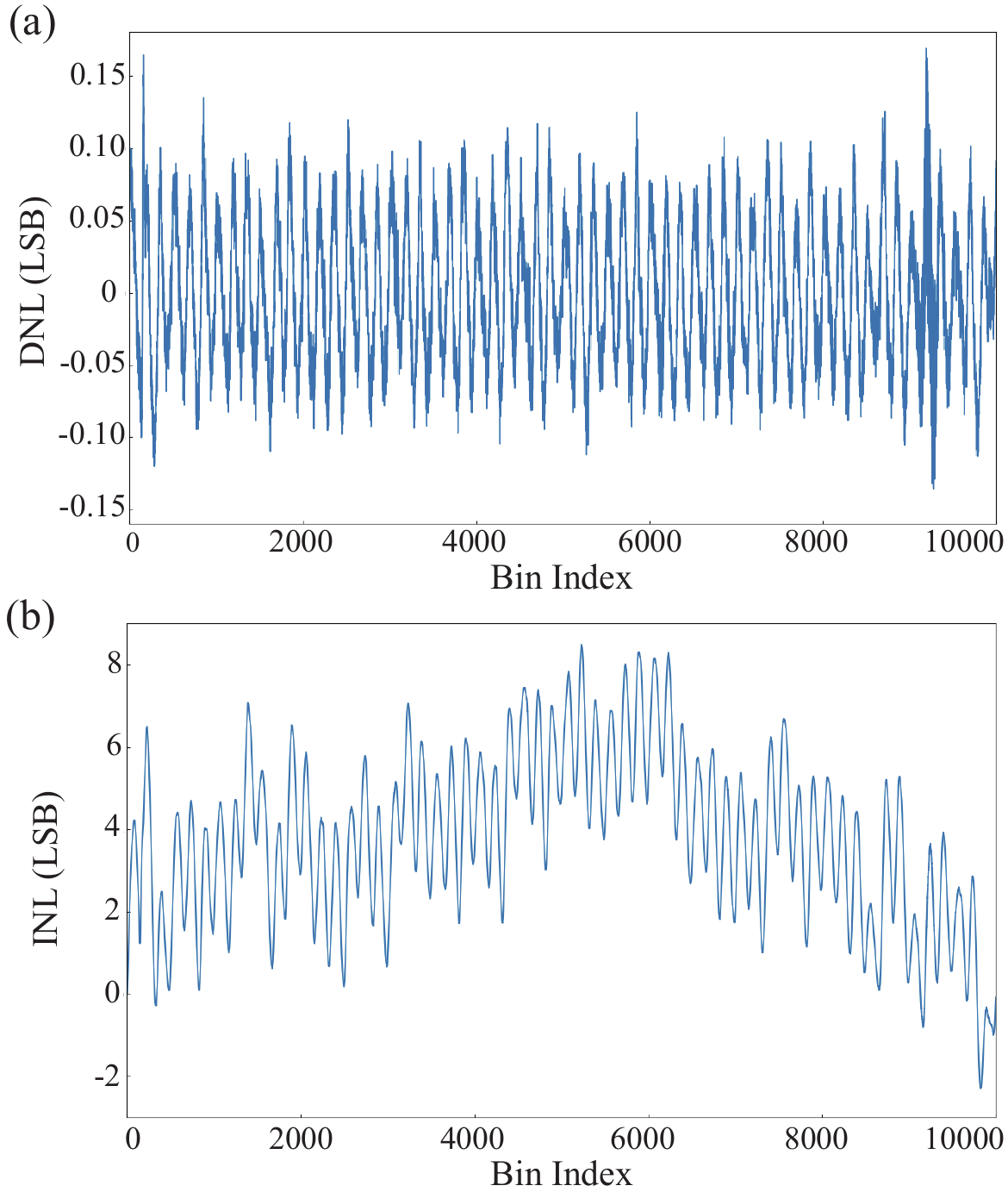}
    \caption{Characterization results of DNL (a) and INL (b). The bin size (LSB) of the TCSPC module is configured as 10 ps.}
    \label{fig3}
\end{figure}

The differential non-linearity (DNL) and integral non-linearity (INL) are significant parameters for the TCSPC module. The DNL is the deviation of the actual bin size from the ideal bin size, i.e., least significant bit (LSB). The INL is calculated as the summation of the DNL from the first bin to the current bin. We follow the method~\cite{DLH84} to characterize the DNL and INL parameters. The results are presented in Fig.~\ref{fig3} (a) and Fig.~\ref{fig3} (b), respectively. The maximum values of the DNL and INL are 0.17 LSB and 8.49 LSB, corresponding to 2 ps and 85 ps, respectively.

\section{NLOS IMAGING EXPERIMENT}

\begin{figure}[htbp]
    \includegraphics[width=8.5 cm]{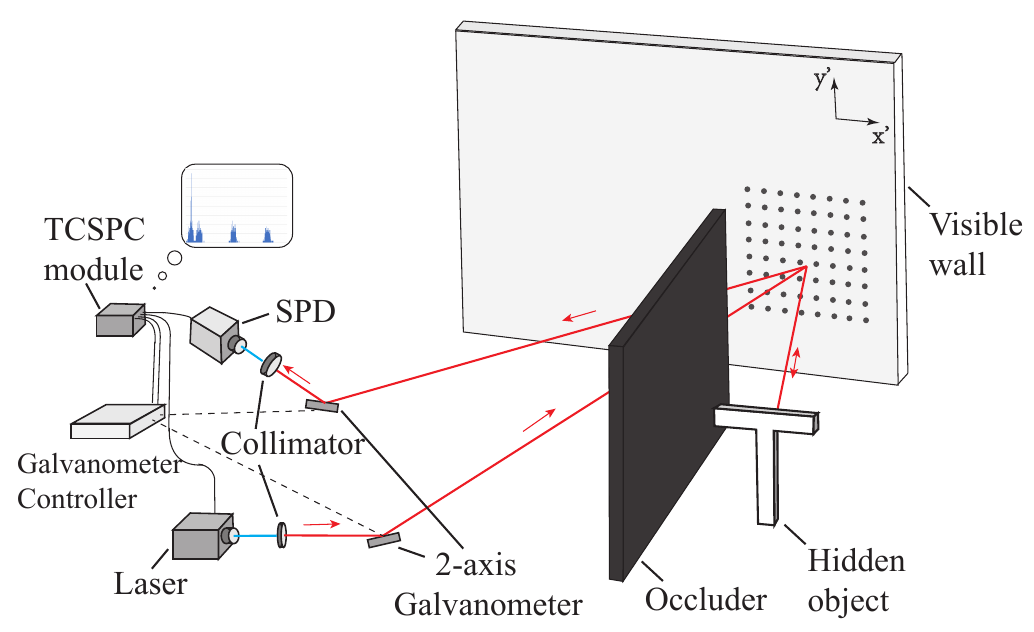}
    \caption{Experimental setup for single-photon TOF NLOS imaging.}
    \label{fig4}
\end{figure}

We then perform a single-photon TOF NLOS imaging experiment using the TCSPC module. The experimental setup is shown in Fig.~\ref{fig4}. A pulsed laser at 1550 nm with 8 ps pulse width and 10 MHz repetition frequency is used to illuminate a small patch of a visible diffuse wall, which is in the direct line of sight. The light scatters off this patch to the hidden object, and then the object reflects parts of the light back to the diffuse wall. In the confocal receiving optical path, the reflected photons from the illuminated patch are collected in a multimode fiber via a collimator and detected by the InGaAs/InP SPD. Two two-axis galvanometers are used to scan an area of 0.8 m $\times$ 0.8 m on the diffuse wall with 64 $\times$ 64 pixels. The signal synchronized to the laser pulses is connected to channel 1 of the TCSPC module as start signal. The SPD output signal is connected to channel 2. The signals generated by the galvanometer controller, which move the directions of pixel $x$ and pixel $y$, are connected to channels 3 and 4, respectively. For each pixel, the TCSPC module creates a TOF histogram. Using all scanning measurement data, a three-dimensional image of the hidden target object is reconstructed following the algorithm in Ref.~\cite{LGM19}.

\begin{figure}[htp]
    \includegraphics[width=8.5 cm]{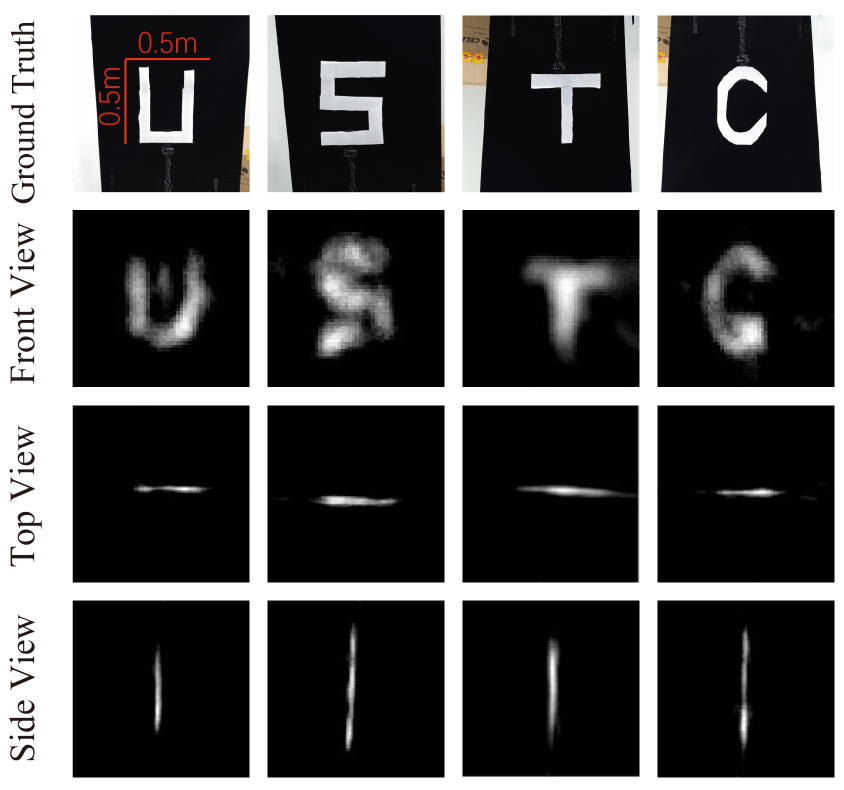}
    \caption{Reconstruction results for the four letter models of ``USTC'' }
    \label{fig5}
\end{figure}

In the experiment, four white plastic letters ``U'', ``S'', ``T'', and ``C'' with a size of $\sim$ 0.5 m are used as hidden objects. The hidden object is $\sim$ 1 m away from visible diffuse wall. The distance between the visible diffuse wall and the SPD is $\sim$ 5 m. As shown in Fig.~\ref{fig5}, the four letters are clearly reconstructed by the proposed scheme. The depth resolution $\delta_{z}$ and lateral resolution $\delta_{x}$ of the imaging system are estimated as 2.3 cm and 6.3 cm, respectively, as calculated by~\cite{WLH21}
\begin{equation}
    \label{zresolution}
    \delta_{z} = c\sqrt{2Ln2(\frac{\sigma cos(\phi)}{2c})^{2}+\frac{T_{SPD}^{2}+T_{laser}^{2}+T_{TCSPC}^{2}}{4}},
\end{equation}
\begin{equation}
    \label{xresolution}
    \delta_{x} = \delta_{z} \frac{\sqrt{w^{2}+z^{2}}}{w}.
\end{equation}
In Eqs.~\ref{zresolution} and ~\ref{xresolution}, $c$ represents the speed of light. $\sigma$ represents the laser diameter, which is 2.67 cm. $\phi$ represents the angle between the hidden object and the visible wall, which is 30$^\circ$. $T_{SPD}$, $T_{laser}$ and $T_{TCSPC}$ represent the FWHM timing jitters of the SPD, laser, and the TCSPC module, which are estimated as 110 ps, 8 ps and 64.5 ps, respectively. $z$ and $w$ represent the distance from the hidden object to the visible wall and half of the scanning width, which are 1 m and 0.4 m, respectively.

From Eqs.~\ref{zresolution} and ~\ref{xresolution}, the spatial resolution of the single-photon TOF NLOS imaging system is primarily contributed by the optical parameters and the SPD timing jitter. Even replacing the TCSPC module with a high-performance commercial TCSPC instrument, e.g., the Time Tagger Ultra with 19 ps FWHM timing jitter (Swabian Instrument)~\cite{timtag}, the depth resolution and lateral resolution can only be slightly improved to 2.2 cm and 5.8 cm, respectively, while the size reaches 18 cm $\times$ 14 cm $\times$ 6 cm. From the perspectives of integration and cost-effectiveness for applications, our TCSPC module is highly suitable for practical single-photon TOF NLOS imaging systems. In addition, the compactness and relatively simple hardware implementation mean that the TCSPC module could be further integrated into the InGaAs/InP SPD. Since the most functions of the TCSPC testing board, such as power, signal preprocessing, logic control, and data transmission, are already exist in the InGaAs/InP SPD, the additional size required for the integration is estimated to be $\sim$4 cm$^{2}$.

\section{CONCLUSION}

In summary, we have implemented a miniaturized TCSPC module dedicated to TOF NLOS imaging applications with a commercially available TDC chip. The TCSPC module exhibits excellent performance with a 10 ps bin size and 27.4 ps minimum RMS time resolution. We have also demonstrated a single-photon TOF NLOS imaging experiment using the TCSPC module, in which a 6.3 cm lateral resolution and 2.3 cm depth resolution are achieved under the conditions of 5 m imaging distance and 1 ms pixel dwell time. Our work provides a practical and cost-effective solution for the development of compact NLOS imaging systems.

\section*{acknowledgments}
This work is supported by the Innovation Program for Quantum Science and Technology under Grant Nos. 2021ZD0300804 and 2021ZD0300300, and the National Natural Science Foundation of China (62175227).

\section*{DATA AVAILABILITY}
Data underlying the results presented in this paper are not publicly available at this time but may be obtained from the authors upon reasonable request.

\par


\end{document}